\documentstyle[twoside,fleqn,npb,epsfig]{article}
\newcommand{\AmS}{{\protect\the\textfont2
  A\kern-.1667em\lower.5ex\hbox{M}\kern-.125emS}}
\newcommand{\msbar}{$\overline{MS}$}
\newcommand{\msba}{\overline{MS}}

\newcommand{\sef}{\sin^2 \theta_{eff}^{lept}}

\newcommand{\sinhat}{\sin^2 \hat{\theta}}

\newcommand{\ini}{\begin{equation}}
\newcommand{\fin}{\end{equation}}
\newcommand{\sms}{\hat{s}^2}
\newcommand{\cms}{\hat{c}^2}
\newcommand{\ems}{\hat{e}^2}
\newcommand{\es}{s_{eff}^2}
\newcommand{\ec}{c_{eff}^2}

\newcommand{\dkc}{\Delta\hat{k}}
\newcommand{\droc}{\Delta\hat{\rho}}
\newcommand{\drcw}{\Delta\hat{r}_W}
\newcommand{\dre}{\Delta r_{eff}}
\newcommand{\dr}{ \Delta r}
\newcommand{\bsli}{\begin{slide}}
\newcommand{\esli}{\end{slide}}
\newcommand{\bcen}{\begin{center}}
\newcommand{\ecen}{\end{center}}

\newcommand{\bite}{\begin{itemize}}
\newcommand{\eite}{\end{itemize}}
\newcommand{\bmath}{\begin{displaymath}}
\newcommand{\dah}{\Delta \alpha_h^{(5)}}

\newcommand{\emath}{\end{displaymath}}
\newcommand{\smallz}{{\scriptscriptstyle Z}} % small letters by paolo
\newcommand{\smallw}{{\scriptscriptstyle W}}

\newcommand{\mz}{M_\smallz}
\newcommand{\mw}{M_\smallw}
\newcommand{\smu}{\hat{s}^2(\mu)}

\title{Topics in Electroweak Physics}
\author{A. Sirlin\address{Department of Physics, 
        New York University, \\
        4 Washington Place, NY 10003, United States}%
        \thanks{Talk presented at the International Symposium Radcor 2002,
        September 8-13, Kloster Banz, Germany.
        This work was supported in part by NSF Grant No.~PHY-0070787.}}
\begin{document}

\begin{abstract}
We briefly discuss five topics in Precision Electroweak Physics: i) the 
recently proposed Effective Scheme of Renormalization, ii) evidence 
for electroweak bosonic corrections derived from the radiative correction 
$\dre$, iii) an approach to estimate the scale of new physics in 
a hypothetical Higgs-less scenario, iv) simple and accurate
formulae for $\es$, $M_W$,
$\Gamma_l$, and their physical applications, v) a recent proposal
concerning the field renormalization constant for unstable particles. 
\end{abstract}

\maketitle

\section{Effective Scheme of Renormalization}

Precise calculations in the Standard Model (SM)
are based on  a number
of renormalization frameworks.
Two of the most frequently employed are:  
1) the On-Shell Scheme (OS) \cite{r1,r2,r3}
2) the \msbar\ approach \cite{r4}. The OS scheme is ``very physical''
in the sense that the renormalized parameters are identified 
with physical, scale-independent observables, such as 
$\alpha$, $G_F$, $\mz$, $\mw$,  \ldots
The \msbar\ approach is frequently applied in a hybrid version, with couplings
defined by \msbar\ subtractions, but retaining physical masses.
It employs scale-dependent parameters such as
$\sms \equiv \sinhat (\mu)$, $\ems (\mu)$ (usually evaluated at $\mu = \mz$) 
and exhibits very good convergence properties \cite{r5}. It
plays an important r\^ole in the analysis of Grand Unified Theories.
However, it leads to a residual scale dependence 
in finite orders of perturbation theory (PT).
Very recently, a novel approach, called the 
Effective Scheme of Renormalization (EFF), was proposed \cite{r6,r7}. 
It shares the good convergence properties of the \msbar\ scheme, but 
it eliminates the residual scale dependence in finite orders of PT. 
A distinguishing feature is that the basic electroweak mixing parameter (EWMP)
is directly identified with $\es \equiv \sef$ , employed by the 
Electroweak Working Group (EWWG) to describe the on-resonance asymmetries
measured at LEP and SLC. It may be evaluated by means of the basic relation
\cite{r6,r7,r8}
\ini \label{eq1}
\es\ \ec = \frac{\pi \alpha}{\sqrt{2}\ G_F M_Z^2 \left( 1- \dre \right)} \ ,
\fin
where  $\dre$ is the relevant radiative correction.
In order to calculate $\dre$ the following strategy was followed:\\
i) Since current calculations of $\es$ incorporate two-loop effects 
enhanced by powers $(M_t^2/M_Z^2)^n$ (with $n=1,2$), we first express 
$\dre$ in terms of corrections $\drcw$, $\droc$, $\dkc$ and $\hat{f}$, 
for which the irreducible 
contributions of this order have been evaluated \cite{r9a,r9}.
ii) To ensure the absence of a residual scale dependence, 
we use scale-independent couplings, such as $e^2$, $\es$, $G_F$, $M_Z^2$,  
retain only two-loop effects enhanced by factors  $(M_t^2/M_Z^2)^n$ ($n=1,2$),
and employ a single definition of the EWMP, identified 
with $\es$.  
In particular, $\sms$ can be expressed in terms of $\es$
by means of the relation
\ini \label{eq2}
\es = \left[  1 + \frac{\hat{e}^2}{\sms} \dkc \left( M_Z^2, \mu \right)\right]\smu \ ,
\fin
where $\dkc(q^2, \mu)$ is an electroweak form factor \cite{r10}.
The analysis leads to the expression \cite{r6}:
\bmath 
\dre  = \drcw - \frac{e^2}{\es} \left[\Delta \hat{\rho} -
\Delta \hat{k} \left( 1-\frac{\es}{\ec}\right)
\right] 
\emath 
\ini  \label{eq3}
-  \frac{e^2}{\es} \ x_t \left[ 2 \Delta \hat{\rho} -
\left( \Delta \hat{\rho} \right)_{lead} - \hat{f} + \Delta \hat{k} 
\frac{\es}{\ec}\right] 
 \ ,
\fin
where $\droc \equiv Re\left[ A_{WW}(M_W^2)
- \cms A_{ZZ}(M_Z^2) \right]/ M_W^2$, 
 $x_t = 3 G_F M_t^2/ (8 \sqrt{2} \pi^2)$, 
$\drcw=-2\delta e/e+(e^2/\sms)\hat{f}$, 
$\left( \Delta \hat{\rho} \right)_{lead} 
= \left( 3 / 64 \pi^2 \right) M_t^2/ M_W^2 $, 
$\hat{f} \equiv \left(Re A_{WW}(M_W^2)-  A_{WW}(0) \right)/ M_W^2 + V_W
+M_W^2 B_W$,
$A_{WW}$ and $A_{ZZ}$ are the $W$ and $Z$ self-energies modulo a factor
$\ems/\sms$,
$V_W$ and $B_W$ are vertex and box corrections contributing to
$\mu$-decay, and $\delta e$ stands for the charge renormalization counterterm.
It is understood that, in Eq.~(\ref{eq3}), $\sms$ is replaced everywhere 
by $\es$. The corrections $\drcw$, $\Delta \hat{\rho}$,
 $\Delta \hat{k}$,  and $\hat{f}$ depend also on $c^2 = M_W^2/M_Z^2$.
In order to obtain an expression that depends solely on $\ec=1-\es$,  
$M_W^2$ is replaced by $c^2 M_Z^2$; in two-loop contributions, 
$c^2$ is replaced by $\ec$, since the difference is of third order;
in one-loop corrections, a Taylor expansion about $c^2 = \ec$ is made,
in conjunction with the one-loop expression for 
$ c^2 - \ec$. The corresponding expression for $M_W$ is given in 
Ref.~\cite{r6}.
An interesting feature is that
the calculation of $\es$ is completely decoupled  from that of $M_W$,
while the  $\es$ results are employed to calculate $M_W$.
The results for the leptonic partial widths $\Gamma_l$ of the $Z$ have been 
recently obtained \cite{r11}. A detailed comparison shows that, for 
$M_H = 100\, \mbox{GeV}$, the difference $|\es(\msba)-\es(EFF)|$ of the \msbar\
and EFF calculation of $\es$ is $\leq 10^{-5}$ over the range
$30\, \mbox{GeV} \leq \mu \leq 200\, \mbox{GeV}$ 
and exhibits a maximum at $\mu \approx 70\, \mbox{GeV}$.
In the $M_W$ case, one finds $|M_W(\msba)-M_W(EFF)| \leq 1\, \mbox{MeV}$
over the range $50\, \mbox{GeV} \leq \mu \leq 205\, \mbox{GeV}$ and
a maximum at $\mu \approx 100\, \mbox{GeV}$. At $\mu = 300\, \mbox{GeV}$,
the differences amount to $\approx 3 \times 10^{-5}$ and $3\, \mbox{MeV}$,
respectively. These findings give support to the usual choice $\mu = M_Z$
in the \msbar\ calculations of observables in the resonance region.
It should be pointed out, however, that this satisfactory state of affairs
holds when the corrections of ${\cal O}\left(\alpha^2 (M_t/M_W)^2 \right)$
are included. If these contributions are excluded, 
$M_W(\msba)$ is a monotonically decreasing function of $\mu$ over the range
$30\, \mbox{GeV} \leq \mu \leq 500\, \mbox{GeV}$ \cite{r12}, and
the choice of scale is very ambiguous. In summary,
the EFF approach has the virtue of eliminating the scale ambiguity which,
in some cases, may create a significant theoretical uncertainty.

\section{Evidence for Electroweak Bosonic Corrections}

It turns out that $\dre$ is very sensitive to electroweak bosonic 
contributions (EWBC), i.e. corrections involving virtual bosons: $W$'s,
$Z$, $H$, $\phi$'s. 
They are subleading numerically, but very important conceptually!
One way to obtain sharp evidence for these corrections is to measure 
$\dre$. Using the current experimental value 
$(\es)_{exp} = 0.23149 \pm 0.00017$ and Eq.(\ref{eq1}),
we find $(\Delta r_{eff})_{exp} = 0.06047 \pm 0.00048$.
On the other hand, subtracting the EWBC, 
the theoretical evaluation leads to: 
 $(\Delta r_{eff})_{theor}^{subtr} = 0.05106 \pm 0.00083$.
The difference is $0.00941 \pm 0.00096$, thus providing evidence 
for the presence of EWBC at the $9.8\  \sigma$ level \cite{r5,r8,r13}!

\section{The Higgs-less Scenario}

The corrections $\dre$ and $\dr$ have been also employed to discuss 
the scale of new physics in a hypothetical scenario in which
the Higgs boson is absent \cite{r14}.
At the one-loop level, the Higgs boson contribution to $\dre$ is 
a complicated function of  $\xi = M_H^2/M_Z^2$, given in Ref.~\cite{r14}.
It may be written in the form
\bmath 
(\dre)_H  = \frac{\alpha}{4 \pi \sms \cms}(\frac{5}{3}-\frac{3}{2}c^2)  
\emath
\ini \label{eq5}
\times \left(\frac{1}{n-4}+ C +\ln{\frac{\mz}{\mu}} \right) + (\dre)_H^{\msba} \, , 
\fin
where the first term is the divergent part and the second one 
is the \msbar-renormalized contribution evaluated at $\mu = M_Z$ 
($C=[\gamma-\ln{4 \pi}]/2$, $\mu = $'t~Hooft scale).
Subtracting $(\dre)_H$ from $\dre$ we have 
\bmath
\dre - (\dre)_H = \dre - (\dre)_H^{\msba} 
\emath
\ini \label{eq6}
- \frac{\alpha}{4 \pi \sms \cms}
(\frac{5}{3}-\frac{3}{2}c^2)(\frac{1}{n-4}+C+\ln{\frac{\mz}{\mu}})\, .
\fin
Clearly, Eq.(\ref{eq6}) is divergent and scale dependent.
We now conjecture that contributions from unknown new physics (NP) cancel
the divergence and scale dependence of Eq.(\ref{eq6}). Thus, the NP
contribution to $\dre$ must be of the form:
\ini
X= \frac{\alpha}{4 \pi \sms \cms} (\frac{5}{3}-\frac{3}{2}c^2) 
\left( \frac{1}{n-4} + C +\ln{\frac{M}{\mu}}\right)  .
\fin
We note that in the \msbar\ renormalization approach, the term
proportional to $\ln{M/\mu}$ represents the NP contribution to $\dre$
at scale $\mu$. If the NP is characterized by a scale $\Lambda$, we may 
decompose
\ini \label{eq8}
\ln{\frac{M}{\mu}} = \ln{\frac{\Lambda}{\mu}} + K\, ,
\fin
where the term involving $K \equiv \ln{\frac{M}{\Lambda}}$ represents the NP 
contribution to  $\dre$ at scale $\Lambda$.
Adding  $X$ to $\dre - (\dre)_H$ we find the expression for 
$\dre$ in the new scenario (NS) in which the Higgs boson contribution has been 
replaced by new physics:
\begin{eqnarray} \nonumber
(\dre)_{NS}& = & \dre - (\dre)_H^{\msba} \\
& + & \frac{\alpha}{4 \pi \sms \cms}
(\frac{5}{3}-\frac{3}{2}c^2) \ln{\frac{M}{M_Z}}\, .
\end{eqnarray}
The last term represents the NP contribution to $\dre$ at scale $M_Z$.
Calculating $\dre - (\dre)_H^{\msba}$ and 
equating $(\dre)_{NS} = (\dre)_{exp}$,
we can determine $\ln{\frac{M}{M_Z}}$. 
Employing $\dah = 0.02761 \pm 0.00036$ and the other experimental inputs,
one finds
\ini \label{eq10}
\ln{\frac{M}{M_Z}} = 0.307 \pm 0.485\, ,
\fin
which corresponds to a central value $M_c = 124\ \mbox{GeV}$ and a 95\% CL
upper bound $M^{95} =  275\ \mbox{GeV}$.
If the model-dependent constant $K$ is positive, we see from Eq.(\ref{eq8})
that $\Lambda$ is sharply bounded: 
$\Lambda \leq  275\ \mbox{GeV}\ @\ \mbox{95\% CL}$.
Instead, if $K < 0$, $\Lambda$ is not bounded by these considerations.
Thus, we can group the NP models into two classes, according to the sign
of $K$. Furthermore, if for instance $\Lambda = 1\ \mbox{TeV}$,
we have  $\ln{\frac{\Lambda}{M_Z}} = 2.395$ and we find from
 Eqs.(\ref{eq8},\ref{eq10}) that $K = -2.088 \pm 0.435$.
Thus, for such $\Lambda$ values, a substantial cancellation of logarithmic 
and constant terms is required \cite{r14,r15}. Similar results are obtained 
from the corresponding analysis of $\dr$ \cite{r14}.

\section{Simple formulae for $\es$, $M_W$, $\Gamma_l$}

Simple formulae that reproduce accurately
the numerical results of the codes in the range
$20\, \mbox{GeV} \le M_H \le 300\, \mbox{GeV}$, probed by
recent experiments,  have been presented \cite{r11}. 
They are of the form:
\begin{eqnarray} \label{eq22a} \nonumber
\es &=&  (\es)_0 + c_1 A_1 + c_5 A_1^2 + c_2 A_2 \\
& &  - c_3 A_3 + c_4 A_4\, , \\
\label{eq22b} \nonumber
M_W &=&  M_W^0 - d_1 A_1 - d_5 A_1^2 - d_2 A_2 \\
& &+ d_3 A_3 - d_4 A_4\, , \\
\label{eq22c} \nonumber
\Gamma_l & =&  \Gamma_l^0 - g_1\,A_1 - g_5\,A_1^2  - g_2\,A_2 \\
& & + g_3\,A_3 - g_4\,A_4 \, , 
\end{eqnarray}
where $A_1 \equiv \ln \left( M_H / 100 \, \mbox{GeV}\right)$,
$A_2 \equiv \left[ \Delta \alpha_h^{(5)}/0.02761 \right]-1$,
$A_3 \equiv \left( M_t / 174.3 \, \mbox{GeV}\right)^2 -1$,
$A_4 \equiv \left[ \alpha_s(M_Z) / 0.118 \right] -1$.
The constants $c_i$, $d_i$, $g_i$ $(i=1-5)$ are given in Ref.~\cite{r11}
for the EFF, \msbar, and OS schemes of renormalization.
Furthermore, Eqs.(\ref{eq22a}-\ref{eq22c}) retain
their accuracy over the range of 
$\dah$ results from recent calculations.
Using Eq.~(\ref{eq22a}) in the EFF scheme and $(\es)_{exp} = $
$ 0.23149 \pm 0.00017$, one finds 
$M_H = 124_{-52}^{+82}\ \mbox{GeV}$ and a 95\% CL upper bound
$M_H^{95} = 280\ \mbox{GeV}$. Instead, Eq.~(\ref{eq22b}) and 
 $(M_W)_{exp} = 80.451 \pm 0.033\ \mbox{GeV}$ lead to
$M_H = 23_{-23}^{+49}\, \mbox{GeV}$,  $M_H^{95} = 122\ \mbox{GeV}$.
Thus, $M_W$ constrains $M_H$ much more sharply than $\es$! It is 
important to note that the  $M_H^{95}$ value derived from $M_W$,
and the direct exclusion bound $M_H > 114\ \mbox{GeV}$ @ 95\% CL,
suggest a very narrow window for $M_H$! 
One may also extract $A_1$ from $(\es)_{exp}$ 
and Eq.~(\ref{eq22a}), to predict $M_W$ via Eq.~(\ref{eq22b}):
$(M_W)_{indir.} = 80.374 \pm 0.025\ \mbox{GeV}$,
which is close to the 
corresponding value $(M_W)_{indir.} = 80.379 \pm 0.023\ \mbox{GeV}$ obtained
in the global analysis \cite{r17}, and differs
from $(M_W)_{exp.}$ by $1.86\ \sigma$.
Finally, we may use simultaneously Eq.~(\ref{eq22a}-\ref{eq22c})
in conjunction with $(\es)_{exp}$, $(M_W)_{exp}$, and $(\Gamma_l)_{exp}$
to obtain $M_H = 97_{-41}^{+66}\ \mbox{GeV}$, $M_H^{95} = 223\ \mbox{GeV}$,
to be compared with  $M_H = 85_{-34}^{+54}\ \mbox{GeV}$, 
$M_H^{95} = 196\ \mbox{GeV}$ in the recent EWWG fit.\\
The current determination of $(\es)_{exp}$
has $\chi^2/\mbox{d.o.f.}=10.6/5$, corresponding to a CL
of only $6\%$, and shows an intriguing
dichotomy: the leptonic observables ($A_l(SLD)$, $A_l(P_\tau)$,
$A_{fb}^{(0,l)}$) lead to  $(\es)_{l}=0.23113 \pm 0.00021$,
while the value from the hadronic ones ($A_{fb}^{(0,b)}$,$A_{fb}^{(0,c)}$,
$<~Q_{fb}~>$) is $(\es)_{h}=0.23220 \pm 0.00029$. 
Thus, there is a $3 \sigma$ difference between the two determinations!
Furthermore, from  $(\es)_{l}$ one finds $M_H = 59_{-29}^{+50}\ \mbox{GeV}$,
$M_H^{95} = 158\ \mbox{GeV}$, closer to the result from $(M_W)_{exp}$.
If $(\es)_{l} - (\es)_{h}$ reflects a statistical fluctuation,
one possibility is  to enlarge the error by $[\chi^2/\mbox{d.o.f.}]^{1/2}$
(PDG prescription), leading to  $\es = 0.23149 \pm 0.00025$.
Interestingly, increasing the error in $\es$ leads to smaller $M_H^{95}$
in the combined $\es$-$M_W$-$\Gamma_l$ analysis: 
$223\ \mbox{GeV} \to 201\ \mbox{GeV}$! The reason is that this procedure
gives enhanced weight to the $M_W$ input, which prefers a smaller $M_H$.
If $(\es)_{l} - (\es)_{h}$ is due to new physics involving the $(t,b)$ 
generation, a substantial,
tree-level change in the $Zb_R\overline{b}_R$ coupling is required \cite{r18}.
If the discrepancy were to settle on the leptonic side, a scenario with
light $\tilde{\nu}$ and $\tilde{g}$ would improve 
the agreement with the 
electroweak data and the direct lower bound on $M_H$ \cite{r19}.\\
It has been pointed out by several people that, if the central values 
of $M_t$ and $M_W$ remain as they are now, but the errors shrink sharply
as expected at Tevatron/LHC or even much better at LC + GigaZ,
a discrepancy would be established with the SM, that can 
be accommodated in the MSSM!\\
The comparison of the calculations of $\es$, $M_W$, and $\Gamma_l$ in 
the EFF, \msbar, and OS frameworks has been applied to study the scheme
dependence and to estimate the theoretical error
arising from the truncation of the perturbative series \cite{r11}.
Including QCD uncertainties, the theoretical errors have been estimated 
to be  $\delta\es \approx 6 \times 10^{-5}$ and 
$\delta M_W \approx 7\ \mbox{MeV}$.

\section{Field Renormalization Constant for Unstable Particles}

In Ref.~\cite{r20} it was found that, in the gauge theory context, the 
conventional definitions of mass and width of unstable particles
are gauge dependent in next-to-next-to-leading order (NNLO).
Furthermore, the conventional expression for the field 
renormalization constant tends to 0 as the particle mass approaches
from below some physical thresholds, which implies the absurd
conclusion that, in such a case, all decays are forbidden.
In Ref.~\cite{r20}, it was proposed that the first problem can be solved 
by considering the complex valued position $\overline{s}$ of the
propagator's pole, which is gauge invariant. We have: 
$\overline{s} = M_o^2 + A(\overline{s})$, where $M_o$ is the bare mass
and $A(s)$ the self-energy. Decomposing 
$\overline{s} = m_2^2 - i m_2 \Gamma_2$, where $m_2$ and $\Gamma_2$ are
real, one identifies $m_2$ and $\Gamma_2$ with the mass and width 
of the particle: 
\ini \label{eq14}
 m_2^2 =  M_o^2 + Re\ A(\overline{s})\, ,
\fin
\ini \label{eq15}
 m_2 \Gamma_2 =  - Im\ A(\overline{s})\, .
\fin
In Ref.\cite{r21}, it was proposed that the second problem can be solved 
by defining the field renormalization constant $\hat{Z}$ by means of
the normalization condition
\ini \label{eq16}
 m_2 \Gamma_2 =  -\hat{Z} Im\ A(m_2^2)\, ,
\fin
which, in conjunction with Eq.~(\ref{eq15}), leads to
\ini \label{eq17}
\hat{Z}^{-1} = 1+\frac{Im \left(A(\overline{s})- A(m_2^2)\right)}{ m_2 \Gamma_2 }\, .
\fin
In the narrow width approximation, the r.h.s. of Eq.~(\ref{eq17})
becomes $1-Re A'(m_2^2) \approx 1-Re A'(M^2)$, 
where $M^2$ is the on-shell mass,
and  $\hat{Z}$ reduces to the conventional expression. 
We note that: i) $\hat{Z}$, defined by  Eq.~(\ref{eq17}), involves
a finite difference, rather than a derivative, thus avoiding the 
threshold problem; ii) using Eq.~(\ref{eq17}) we see that the r.h.s.
of  Eq.~(\ref{eq16}) is gauge invariant, since it equals  $m_2 \Gamma_2$
as a mathematical identity. It was also shown that the use of  
Eq.~(\ref{eq17}) removes unphysical threshold singularities in
the relation between on-shell and pole widths \cite{r22}.
This approach has been recently discussed in the framework of 
renormalization theory \cite{r23,r24}. 
Dividing the unrenormalized transverse propagator 
$-i Q_{\mu \nu} [s - M_o^2 - A(s)]^{-1}$ 
($ Q_{\mu \nu} = g_{\mu \nu} - q_\mu q_\nu / q^2$) by  $\hat{Z}$,
and introducing $S(s) \equiv \hat{Z} A(s)$, 
$\delta M^2 \equiv Re S(\overline{s})$, $\hat{Z}\equiv 1-\delta\hat{Z}$,
we obtain the renormalized propagator:
\ini \label{eq18}
{\cal D} = -i Q_{\mu \nu} / (s - m_2^2 - S^{(r)}(s))\, ,
\fin
where
\ini \label{eq19}
 S^{(r)}(s) = S(s)-\delta M^2 +\delta\hat{Z}(s-m^2_2)
\fin
stands for the renormalized self-energy. Since $\delta M^2$ and 
$\delta\hat{Z}$ are real, they should be chosen so that $Re\ S^{(r)}(s)$
is ultraviolet convergent to all orders. Once this is done, 
$Im\  S^{(r)}(s) = Im\ S(s) = \hat{Z}\ Im\ A(s)$ must also be convergent, 
since there are no additional counterterms available. This means 
that  $\hat{Z}$ may be defined by imposing an appropriate 
normalization condition on $Im\ S(s)$. A particularly simple one is
\ini
Im S(m_2^2) = - m_2 \Gamma_2\, ,
\fin
which coincides with Eq.~(\ref{eq16})!
It was proposed independently in Ref.~\cite{r21} to solve the threshold 
and gauge-dependence problems, and in  Ref.~\cite{r24} to implement a 
systematic order by order removal of the ultraviolet divergences in 
$S^{(r)}(s)$. In Ref.~\cite{r23} it was also emphasized 
that in this formulation one can derive closed and exact expressions for
the mass and field-renormalization counterterms, to wit 
\ini \label{eq21}
\delta M^2 =  Re S(\overline{s})\, ;\, 
\delta\hat{Z}=\frac{Im \left[S(\overline{s})-S(m_2^2)\right]}
{m_2 \Gamma_2}\, . 
\fin
In many cases, $\Gamma_2 = {\cal O}(g^2)$, where $g$ is a generic 
gauge coupling. If $\delta M^2$ and $\delta\hat{Z}$ admit expansions in
powers of $\Gamma_2$, they can be expressed as series involving 
$R \equiv Re\ S(s)$, $I \equiv Im\ S(s)$, and their powers and derivatives 
evaluated at $s = m_2^2$. These expansions of Eq.~(\ref{eq21})
coincide with the order by order analysis in Ref.~\cite{r24}. 
However, in other 
important instances, such as the photonic corrections to the $W$
self-energy, such expansions are ill-defined and lead to power-like
infrared divergences! In such cases one should employ the exact formulae
in Eq.~(\ref{eq21}), which lead to sensible expressions for 
 $\delta M^2$, $\delta\hat{Z}$, and the renormalized propagator \cite{r23}.\\

%\section*{Acknowledgments}

\end{document}